\documentclass[letterpaper, 10 pt, conference]{ieeeconf}  
\IEEEoverridecommandlockouts                             
\usepackage{amsmath,amssymb,amsfonts}
\usepackage{cases}
\usepackage{graphicx}
\usepackage{subcaption}
\usepackage{tikz}
\usepackage{cite}
\usepackage{pgfplots, pgfplotstable}
\pgfplotsset{compat=newest}
\usetikzlibrary{shapes,positioning,calc}
\usetikzlibrary{decorations.markings}

\newcommand{\qu}{\overline{q}}
\newcommand{\ql}{\underline{q}}
\newcommand{\pu}{\overline{p}}
\newcommand{\pl}{\underline{p}}
\newcommand{\uu}{\overline{u}}
\newcommand{\ul}{\underline{u}}

\newcommand{\rootvertice}{0} 
\newcommand{\G}{\mathcal{G}}
\newcommand{\V}{\mathcal{V}}
\newcommand{\E}{\mathcal{E}}
\newcommand{\Path}{\mathcal{P}}

\newtheorem{theorem}{Theorem}

\newtheorem{remark}{Remark}

\title{\LARGE \bf
On properties of hydraulic equilibria in district heating networks
}

\author{Ask Hällström$^{1,\dagger,\star}$, Felix Agner$^{1, \ddagger}$, Richard Pates$^{1,\dagger,\ddagger}$
\thanks{$^{\star}$Corresponding author: \tt \small{ask.hallstrom@control.lth.se}}
\thanks{$^{1}$Department of Automatic Control, Lund University, Sweden. The authors are members of the ELLIIT Strategic Research Area at Lund University.}%
\thanks{$^{\dagger}$The authors are grateful to the initiative COMPEL, with funding from the Swedish government.}%
\thanks{$^\ddagger$This work is funded by the European Research Council (ERC) under the European Union's Horizon 2020 research and innovation program under grant agreement No 834142 (ScalableControl).}%
}

\usepackage[hidelinks]{hyperref}
\usepackage[capitalize]{cleveref}

\begin{document}

\maketitle
\thispagestyle{empty}
\pagestyle{empty}

\begin{abstract}
District heating networks are an integral part of the energy system in many countries. In future smart energy systems, they are expected to enhance energy flexibility and support the integration of renewable and waste energy sources. An important aspect of these networks is the control of flow rates, which dictates the heat delivered to consumers. This paper concerns the properties of flow rates in tree-structured district heating networks. We show that under mild assumptions of monotonicity in the hydraulic network components, statements regarding the stationary flow rate distribution can be made. In particular, when all consumers in a network incrementally open their valves, an increase in total flow rate throughput is guaranteed, while if one consumer does not open their valve when others do, they will receive a reduced flow rate. These properties are illustrated numerically on a small 2-consumer network as well as on a larger 22-consumer network. Previous works have shown that these properties allow the design and use of efficient control strategies for optimal heat distribution.

\end{abstract}

\section{INTRODUCTION} \label{sec:introduction}

District heating networks are considered a key enabler of future smart energy systems \cite{lund_smart_2017}, facilitating the integration of renewable and waste energy sources into the energy mix \cite{lund_4g_2018}. The transition to modern district heating networks brings with it many control-related challenges, such as intermittent heat supply and demand, interconnections of diverse heat sources, and numerous actuators such as pumps and valves \cite{vandermeulen_controlling_2018}. However, there are also many new opportunities for designers of these control systems, such as increased metering and monitoring, smarter actuators, and greater communication possibilities \cite{vandermeulen_controlling_2018}. This has sparked several new research trends within the control systems community, in particular focusing on exploitable properties of district heating models. The present paper concerns useful equilibrium properties of district heating hydraulics in tree-structured networks.

The dynamic properties of district heating hydraulics have been the topic of several recent papers. In \cite{de_persis_proportional_2008}, pressure regulation was considered in a small 2-consumer network and was subsequently extended to proportional control of larger networks in \cite{de_persis_pressure_2011} and proportional-integral control in \cite{de_persis_output_2014}. In \cite{hauschild_port_hamiltonian_2020,strehle_port-hamiltonian_2022}, district heating hydraulic models are considered from a port-Hamiltonian perspective, and in \cite{strehle_port-hamiltonian_2022,machado_modeling_2022,strehle_unifying_2024}, passivity properties of dynamic hydraulics are shown for varying structures of district heating networks.

These previous works show how the dynamical properties of district heating hydraulics are well behaved and can be controlled with smart and simple control-schemes. Moving from flow control to heat distribution also requires an understanding of the thermodynamics in both the network and the connected buildings. When modeling this, a common assumption is that the hydraulic dynamics are fast in comparison to the slower thermodynamics. Thus static hydraulic relations can be employed. This assumption is valid when the controllers that govern the system thermodynamics utilize higher-level control using flow rates as a control input, which in turn is actuated by a lower-level hydraulic control layer. Such a setup is considered in e.g., \cite{agner_bottlenecks,ahmed_control-oriented_2023,machado_decentralized_2023,sibeijn_dissipativity_2024}. It is also valid when the slower signals in the thermodynamic domain are used to govern the hydraulic actuators, as is typically the case in existing district heating networks \cite[p.470]{thebible}. In these scenarios, understanding the static relations between hydraulic actuators and the resulting network flow rates (which in turn dictate the distribution of heat) is important. Such static hydraulic models for district heating networks were considered in \cite{jeeninga_existence_2023}. There, the existence of equilibria in networks with large numbers of actuators was considered.

The present paper extends previous results by considering properties of hydraulic equilibria in a particular class of networks. We consider tree-structured, single-producer district heating networks in which the control of flow rates is actuated through valves. This valve-actuated flow rate control scheme is the norm in existing networks \cite[p.470]{thebible}, and while meshed network topologies are increasingly common, we focus on tree-structured networks in this work for two reasons. First, they remain in use, especially in smaller or older systems, and second, the directed structure allows us to derive rigorous properties that would be harder to prove in the meshed setting. Thus, the presented results should be viewed as a first step toward understanding more general topologies. In this type of tree-structured network an interesting problem can arise under peak cold conditions, where the hydraulic limitations of the network may not allow the delivery of sufficient heat to all consumers \cite{agner_bottlenecks}. In this scenario, understanding hydraulic equilibrium properties is key for enabling efficient heat distribution. 

For the considered class of networks, we establish properties of the steady-state map from hydraulic actuators (valve positions) to the flow rates delivered to consumers. We prove two key input-output properties: First, if all control valves are incrementally opened, then the total flow rate throughput to the consumers increases. Second, if one control valve remains fixed while others are incrementally opened, the flow rate through said valve decreases. We prove these statements only assuming incremental monotonicity of the functions that describe the hydraulic network components. This potentially allows the results to be applied to other types of networked systems. Previous works \cite{agner_automatica_preprint} have shown how the aforementioned equilibrium properties as proven in this paper, though deceptively simple, enable design of scalable control strategies for optimal heat distribution.

The manuscript is structured as follows. We begin by presenting a mathematical model of steady-state district heating network hydraulics in \cref{sec:modeling}. We then present and prove the two aforementioned input-output properties in \cref{sec:main results}. These theoretical results are validated in numerical experiments in \cref{sec:experiments}. Finally, in \cref{sec:conclusion}, the paper is concluded with discussion and remarks on open research directions.

\section{MATHEMATICAL MODELING OF DISTRICT HEATING NETWORKS} \label{sec:modeling}
This section details the models we employ to analyze hydraulic properties of district heating networks. Note that the present paper concerns stationary hydraulic models. For dynamical models and properties of district heating network hydraulics, we refer to \cite{strehle_unifying_2024}.

\subsection{Graph representation} \label{subsec:graph model}
We model district heating networks by directed graphs $\G = (\V, \E)$, where $\V$ denotes the set of vertices and $\E$ the set of edges. The networks consist of two topologically symmetrical layers; a supply-side and a return-side. This assumption is reasonable, as supply and return pipes are often installed together, typically within the same conduits or trenches. For a small example see \cref{fig:2-consumer network}. In our model, each edge $(i,j) \in \E$ is hence associated with two pipes, one in the supply-layer and one in the return-layer. We associate with each such pair supply-and-return-flow rates $q_{ij}^s$ and $q_{ij}^r$, where the supply-side flow $q_{ij}^s$ goes from $i$ to $j$ and the return-side flow $q_{ij}^r$ goes reversely from $j$ to $i$.

Each vertex $i \in \V$ corresponds to two topologically symmetrical points in the supply-and-return networks, $i^s$ and $i^r$ respectively. We denote the pressure in these points $p_i^s$ and $p_i^r$, and the differential pressure $p_i = p_i^s - p_i^r$. In each vertex, we denote the flow rate $q_i$ as the flow from the supply-side network to the return-side network. 
When a vertex does not represent connections between the supply-and-return sides, such as pumps or valves, we simply set $q_i=0$.

\subsection{Hydraulic components} \label{subsec:components}
We model the network with three types of components: pipes, valves and pumps. Pump vertices $i$ are modeled as ideal sources of pressure
$$p_i^s-p_i^r = \text{const.}$$
Pumps are typically operated at large facilities, such as factories with recoverable waste heat, or dedicated heating production plants.

Pipes, which correspond to edges $(i,j) \in \E$, are commonly modeled through the Darcy-Weisbach equation in a quadratic relationship: \cite[p.442]{thebible}
$p_i^s-p_j^s=k_{ij}^sq_{ij}^s\left|q_{ij}^s\right|$ for the supply network and $p_j^r-p_i^r=k_{ij}^rq_{ij}^r\left|q_{ij}^r\right|$ for the return network. $k_{ij}^s$ and $k_{ij}^r$ are pipe-specific constants which depend, among other things, on the turbulence induced by the pipe geometry. The results of this paper do not rely specifically on the Darcy-Weisbach model, and hence we generalize this relation to
$$p_i^s-p_j^s=f_{ij}^s\left(q_{ij}^s\right)$$
$$p_j^r-p_i^r=f_{ij}^r\left(q_{ij}^r\right)$$
where the functions $f_{ij}^s$ and $f_{ij}^r$ are strictly increasing, corresponding to the friction-induced pressure losses due to increased flow rates.

Finally, the network includes consumers. We model the consumers as valves, which in turn are typically modeled through a manufacturer-provided valve curve. An example of a relation for a linear valve curve is $p_l^s-p_l^r=k_lu_l^{-2}q_l\left|q_l\right|$, where $q_l$ is the flow through the valve, $k_l$ is a constant and $u_l$ is the valve position. Here we will again generalize this model as
$$p_l^s-p_l^r=g_l(q_l,u_l),$$
where $g_l$ is strictly increasing in the first argument and strictly decreasing in the second for $u_l$ in the normal operating range. Note that consumer substations typically consist also of a heat exchanger and piping. For the purpose of this paper, we lump these resistive components of the consumer substation together with the supply and return pipes which constitute the network.

\subsection{Network properties} \label{subsec:network}
In this work, we assume that the network is tree-structured. We assume that the root vertex of the tree, denoted $\rootvertice$, is a pump and we assume that the leaf vertices of the tree, $\V_L$, represent valves (i.e., they are consumers). Each vertex in the graph which is neither the root nor a leaf vertex represents a network junction. 

Under this structure, assuming no leakage and taking the incompressibility of water into account, we can observe a symmetry between the flow rates in the supply-and-return lines: $q_{ij}^s = q_{ij}^r = q_{ij}$. This holds since we are considering a closed hydraulic circuit, with no loops in supply or return networks. Hence if $q_{ij}^s \neq q_{ij}^r$, the total volume of water on either side of the edge would increase or decrease. The network also admits a relationship analogous to Kirchhoff's current law in electrical systems: the flow entering a junction in either the supply or return network must equal the flow exiting the junction. This fundamental relation is essential for much of the analysis presented in this paper. By considering a vertex $j$ in either of the networks we find
$$q_j = \sum_{j: (i,j) \in \E}q_{ij}-\sum_{k:(j,k)\in \E}q_{jk}.$$

To simplify the subsequent theory, we will also consider equations for the differential pressure $p_i$, now without the superscript. For the pump located at the root vertex $\rootvertice$, we now have
$$p_\rootvertice = \text{const}.$$
For leaf vertices, with consumer valves, we have
$$p_l = g_l(q_l,u_l).$$
For the pipes we can find a similar relation by considering
$$p_i-p_j = \left(p_i^s-p_i^r\right)-\left(p_j^s-p_j^r\right)$$
$$=\left(p_i^s-p_j^s\right)+\left(p_j^r-p_i^r\right) = f_{ij}^s(q_{ij})+f_{ij}^r(q_{ij}).$$
By denoting the sum $f_{ij}(q_{ij})=f_{ij}^s(q_{ij})+f_{ij}^r(q_{ij})$ we find the simple relation
$$p_i-p_j=f_{ij}\left(q_{ij}\right),$$
where $f_{ij}$ is also a strictly increasing function.

\subsection{Illustrative example} \label{subsec:example}
\begin{figure}[h]
    \centering
    \includegraphics[]{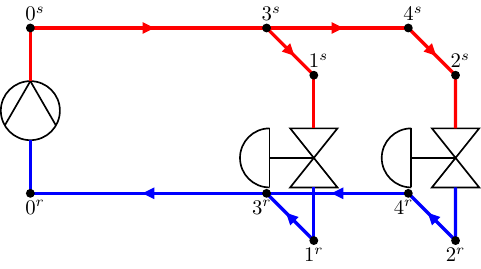}
    \caption{Small two-consumer network with one pump and two valves.}
    \label{fig:2-consumer network}
\end{figure}
To illustrate these relations, we construct the hydraulic equilibrium equations for the small 2-consumer example seen in Figure \ref{fig:2-consumer network}. We model the pipes with the Darcy-Weisbach equation ($k_{ij}^s = k_{ij}^r = 0.5$), the valves with a linear valve curve ($k_l = 1$) and set $p_\rootvertice = 1$ for the pump.

From this the differential pressures of the system are given by the flows according to
$$\begin{cases}
    p_0-p_3=q_{03}|q_{03}|,\\
    p_3-p_4=q_{34}|q_{34}|,\\
    p_3-p_1=q_{31}|q_{31}|,\\
    p_4-p_2=q_{42}|q_{42}|.\\
\end{cases}$$
From the conservation of flow we also have
$$\begin{cases}
    q_0=-q_{03},\\
    q_1=q_{31},\\
    q_2=q_{42},\\
    q_3=0=q_{31}+q_{34}-q_{03},\\
    q_4=0=q_{42}-q_{34},\\
\end{cases}$$
and finally for the valves we have 
$$\begin{cases}
    p_1 = u_1^{-2}q_1|q_1|,\\
    p_2 = u_2^{-2}q_2|q_2|.
\end{cases}$$

\section{MAIN RESULTS} \label{sec:main results}
In this section we present a theorem which, when applied to heating networks, describes how the flows to consumers are affected by changes in valve positions. The theorem shows that in tree-structured networks with a single pump, increasing pump pressure or opening valves always leads to an increase in the total flow reaching consumers, under the monotonicity conditions of pipe and valve functions described in \cref{sec:modeling}. Conversely, if some valves are opened while pump pressure remains constant, consumers with unopened valves will experience a decrease in flow.

Motivated by the modeling in Section II, we consider equations constructed from graphs. From a directed, rooted tree graph $\G=(\V,\E)$ with vertices $\V$ denoted by integers, and edges $\E$, we can denote its root with $0$ and the set of its leaves by $\V_L\subset\V$. If a function $f_{ij}$ is given for each edge $(i,j)\in\E$ and a function $g_l$ is given for each leaf vertex $l\in\V_L$, we can define a set of equations constructed from the functions and the graph as
\begin{subnumcases}{\label{eq:theorem conditions}}
p_l = g_l(q_l,u_l) \quad \forall l \in \V_L,\\
p_i-p_j = f_{ij}(q_{ij}) \quad  \forall (i,j) \in \E,\\
q_j = 0 \quad  \forall j \in \V \setminus \left(\V_L \cup \{ \rootvertice \}\right),\\
q_j = \sum_{i: (i,j) \in \E}q_{ij}-\sum_{k:(j,k)\in \E}q_{jk} \quad  \forall j \in \V.
\end{subnumcases}

A tuple of vectors $(p^*,q^*,u^*)$ is considered a solution if, upon substituting $p_j = p^*_j$ and $q_j = q^*_j$ for all vertices $j \in \V$, $q_{ij} = q^*_{ij}$ for all edges $(i,j) \in \E$, and $u_l = u^*_l$ for all leaf vertices $l \in \V_L$, the equations \eqref{eq:theorem conditions} are satisfied.

\begin{theorem}\label{thm: main result}
Consider the set of equations \eqref{eq:theorem conditions} defined by a given directed, rooted tree graph $\G=(\V,\E)$ where the functions $f_{ij}$ are strictly increasing and the functions $g_l$ are strictly increasing in the first argument and strictly decreasing in the second. Suppose that $(\pu,\qu,\uu)$ and $(\pl,\ql,\ul)$ are two solutions to the equations with $\pu_0\geq\pl_0$. If for all $l\in\V_L$ we have $\uu_l\geq\ul_l$ then it holds that
\begin{equation} {\label{eq:result 1}}
    \sum_{l \in \V_L} \qu_l \geq \sum_{l \in \V_L} \ql_l,
\end{equation}
with a strict inequality if $\pu_0>\pl_0$ or there exists an $l\in\V_L$ such that $\uu_l>\ul_l$.

Furthermore, assume that the root vertex of $\G$ has out-degree one, that $\pu_0=\pl_0$, and that there exists an $l\in\V_L$ such that $\uu_l>\ul_l$. If there also exists a $k\in\V_L$ such that $\uu_k=\ul_k$, it holds that
\begin{equation} {\label{eq:result 2}}
    \qu_k < \ql_k.
\end{equation}

\end{theorem}
\vspace{0.2cm}
\begin{remark}
    In the context of district heating networks the pump, which corresponds to the root vertex, is in practice always connected in series with a pipe. This gives the root vertex out-degree of one.
\end{remark}
\newpage

\begin{proof}
To prove Theorem 1 we begin by stating some useful properties of all solutions to \eqref{eq:theorem conditions}. Using (1c) and the tree-structure of $\G$, (1d) can be split into three cases
\begin{subnumcases} {\label{eq: flow}}
q_\rootvertice = -\sum_{i:(\rootvertice,i)\in\E}q_{\rootvertice i},\\
q_{ij} = \sum_{k:(j,k)\in\E}q_{jk} \quad \forall (i,j)\in\E : j \in \V \setminus \left(\V_L \cap \{ \rootvertice \}\right),\\
q_{kl} = q_l \quad \forall (k,l)\in\E : l\in\V_L.
\end{subnumcases}
Furthermore, summing over all vertices and using (1c) and (1d), we obtain: 
$$\sum_{j\in\V}q_j =\sum_{j\in\V}\left(\sum_{i: (i,j)\in\E}q_{ij}-\sum_{k: (j,k)\in\E}q_{jk}\right)=0\implies$$
\begin{equation}\label{eq: total flow}
    q_\rootvertice+\sum_{l\in\V_L}q_l=\sum_{j\in\V}q_j = 0
\end{equation}
Using these properties we now prove the result in \eqref{eq:result 1}, by contradiction. To this end: Assume that there exists two solutions, as in the theorem statement, with $\pu_0\geq\pl_0$ and $\forall l \in \V_L : \uu_l \geq \ul_l$ , but where $\sum_{l\in\V_L}\qu_l < \sum_{l\in\V_L}\ql_l$.

We will now show that this assumption leads to a path from the root to a leaf in the graph, where $\qu_{ij}<\ql_{ij}$, and $\pu_i\geq\pl_i$ holds along the path. Upon reaching the leaf vertex these inequalities are what results in a contradiction.

To do so we first note that \eqref{eq: total flow} gives us $-\qu_0<-\ql_0$ which can be rewritten by (4a) into $\sum_{i:(0,i)\in\E} \qu_{0i}<\sum_{i:(0,i)\in\E} \ql_{0i}$. Since the first sum is smaller than the second and they contain the same number of elements, we are guaranteed to find at least one element in the first sum that is smaller than the corresponding element in the second. Hence we must have $\qu_{0i}<\ql_{0i}$ for some child of the root $i$.

Furthermore, for any edge $(i,j)\in\E$ we find that if $\qu_{ij}<\ql_{ij}$ and $\pu_i\geq\pl_i$ we have from (1b) and the strictly increasing property of the functions $f_{ij}$ that $\pu_j>\pl_j$ which of course also implies $ \pu_j\geq\pl_j$. If vertex $j$ is not a leaf we also have from (4b) that $\sum_{k:(j,k)\in\E} \qu_{jk}<\sum_{k:(j,k)\in\E} \ql_{jk}$ and hence there must be at least one child $k$ of $j$ with $\qu_{jk}<\ql_{jk}$.

From any edge $(i,j)\in\E$ where $\qu_{ij}<\ql_{ij}$ and where $\pu_i\geq\pl_i$ we can now iterate the argument above to find a path $\Path=((i,j),\dots,(k,l))$ from $i$ to a leaf vertex $l$, where $\qu_{jk}<\ql_{jk}$ for all $(j,k)\in\Path$ and where $\pu_j\geq\pl_j$ holds for all vertices $j$ in the path.

Upon reaching a leaf vertex $l$, the existence of such a path would however lead to a contradiction. Since the leaf vertex $l$ is part of the path we have $\pu_l\geq\pl_l$, but using (4c) on the leaf vertex gives us $\qu_l<\ql_l$ which together with the assumption of $\forall l\in\V_L:\uu_l\geq\ul_l$ and the properties of the functions $g_l$ results in $\pu_l = g_l(\qu_l,\uu_l) < g_l(\ql_l,\ul_l) = \pl_l$ from (1a).

Starting from the root we have a path $\Path$ with the properties described above and the implied contradiction thereby proves the inequality in \eqref{eq:result 1}.

To also prove the strict inequality, we will now show how assuming an equality leads to a contradiction under either of the two new stronger assumptions described in the theorem. Namely when \( \pu_0 > \pl_0 \) or when there exists an \( l \in \V_L \) such that \( \uu_l > \ul_l \).

We again assume that $\pu_0\geq\pl_0$ and $\forall l \in \V_L : \uu_l \geq \ul_l$ but now consider the case where $\sum_{l\in\V_L}\qu_l = \sum_{l\in\V_L}\ql_l$.

This gives us from \eqref{eq: total flow} that $\qu_0=\ql_0$ and from (4a) that $\sum_{i:(0,i)\in\E}\qu_{0i} = \sum_{i:(0,i)\in\E}\ql_{0i}$. If two sums, with the same number of elements are equal we must either have that all elements are equal, or that at least one element is smaller in one sum than the corresponding element in the other. Hence we can either find a child vertex $i$ to the root so that $\qu_{0i}<\ql_{0i}$, in which case we can form a path as before and reach a contradiction, or we have that $\qu_{0i}=\ql_{0i}$ for all children $i$ of the root.

If $\qu_{ij}=\ql_{ij}$ and $\pu_i\geq\pl_i$ for any edge $(i,j)\in\E$ we have from (1b) and the strictly increasing property of the functions $f_{ij}$ that $\pu_j\geq\pl_j$. If vertex $j$ is not a leaf we also have from (4b) that $\sum_{k:(j,k)\in\E} \qu_{jk}=\sum_{k:(j,k)\in\E} \ql_{jk}$ and now $\qu_{jk}=\ql_{jk}$ must hold for all children $k$ of $j$ otherwise there would be a child $k$ to $j$ so that $\qu_{jk}<\ql_{jk}$ in which case the relation $\pu_j\geq\pl_j$ would allows us to form a path as before, resulting in a contradiction.

We can now iterate from the root to show that $\qu_{ij}=\ql_{ij}$ must hold for all edges $(i,j)\in\E$ and from that we can now consider the two cases described in the theorem.

Case 1: If $\pu_0>\pl_0$ we can choose any path from the root vertex to a leaf vertex and conclude that $\pu_i>\pl_i$ for all vertices in the path using (1b). For the leaf vertex $l$ at the end of the path we then get from (1a) that $g_l(\qu_l,\uu_l)=\pu_l>\pl_l = g_l(\ql_l,\ul_l)$ which is a contradiction since $\qu_l=\ql_l$ and $g_l$ is decreasing in the second argument with $\uu_l\geq\ul_l$.

Case 2: If instead $\uu_l > \ul_l$ for some $l\in\V_L$ we can follow the path from the root to vertex $l$ and conclude that since $\pu_0\geq\pl_0$ we must have $\pu_j\geq\pl_j$ for all vertices $j$ along the path. When we reach the leaf $l$ we have from (4c) that $\qu_l=\ql_l$ and from this and the properties of $g_l$ we find that $\pu_l = g_l(\qu_l,\uu_l) < g_l(\ql_l,\ul_l)=\pl_l$ according to (1a). This is a contradiction since we had $\pu_l\geq\pl_l$ from the path.

Together, these two cases prove that having $\pu_0>\pl_0$ or $\exists l\in\V_L : \uu_l > \ul_l$ guarantees a strict inequality in \eqref{eq:result 1}.

Finally, to prove the last part of the theorem we begin by noting that solutions to the equations in \eqref{eq:theorem conditions} for a graph $\G$ can be used to find solutions also for the equations applied to subtrees of $\G$. If $j$ is the vertex of $\G$ that serves as new root for a subtree, and $i$ is its parent, then the corresponding subset of the solutions to (1) for $\G$ will also be solutions for (1) on the subtree if $q_j^*$ in the solutions is replaced by $-q_{ij}^*$.

Using this fact we now prove the result in (3), again by contradiction. To this end:
Assume that we have a directed, rooted tree graph $\G$ where the root has a single child, and again two solutions to \eqref{eq:theorem conditions}, as in the theorem statement, with $\pu_0=\pl_0$, $\forall l \in \V_L:\uu_l \geq \ul_l$, and $\exists l\in\V_L:\uu_l>\ul_l$ but where $\qu_k\geq \ql_k$ for some $k\in\V_L$ where $\uu_k=\ul_k$.

We will now show, using the first part of the theorem on larger and larger subtrees of $\G$, that this assumption leads to a path from the leaf node $k$ to the child of the root where $\pu_j\geq\pl_j$ for all vertices $j$ on the path. Using the first part of the theorem on the subtree formed from the child of the root will then, using the out-degree of one condition, result in a contradiction.

Using (1a) and the properties of $g_k$, we have that $\pu_k\geq \pl_k$. Since $k$ is a leaf vertex we have from (4c) that $q_k=q_{jk}$ where $j$ is the parent of $k$, hence we have $\qu_{jk}\geq\ql_{jk}$ and by (1b) we also have $\pu_j\geq \pl_j$.

Now consider a subtree of $\G$ with its root in an arbitrary vertex $j$ with parent $i$, and with $\pu_j\geq \pl_j$. From the solutions of \eqref{eq:theorem conditions} formed by $\G$ we have that for all leaf vertices $l$ it holds that $\uu_l\geq\ul_l$ and hence $\uu_l\geq\ul_l$ must hold also for all leaf vertices $l$ in the subtree. By using \eqref{eq:result 1}, i.e. the first part of the theorem statement together with \eqref{eq: total flow} on the root of the subtree we find that $\qu_{ij}\geq\ql_{ij}$, since $\qu_{ij}$ and $\ql_{ij}$ act as $-\qu_{0}$ and $-\ql_{0}$ for the subtree. Hence by (1b) we see that starting with $\pu_j\geq\pl_j$ also gives us $\pu_i\geq\pl_i$.

By iterating this we can now find a path starting in vertex $k\in\V_L$, where $\uu_k=\ul_k$, to the only child of the root $i$, where $\pu_j\geq \pl_j$ holds for all vertices $j$ in the path.

In the final step we conclude that if $i$ denotes the only child of the root, we have $\pu_i\geq \pl_i$ and from the assumption also that $\exists l\in\V_L:\uu_l>\ul_l$, where $l$ is guaranteed to be one of the descendant of $i$ since the out-degree of the root is one. Hence applying the first part of the theorem on the subtree with its root in $i$ gives a strict inequality in (2) and using (5) on the new root we get $\qu_{0i}>\ql_{0i}$. From $\pu_i\geq\pl_i$, the properties of $f_{0i}$ and (1b) we now reach the root with $\pu_0 > \pl_0$ which is a contradiction since we assumed $\pu_0=\pl_0$. Hence the result in \eqref{eq:result 2} must also hold.

\end{proof}
\section{NUMERICAL EXPERIMENTS} \label{sec:experiments}

To visualize and investigate the theoretical results in a more practical setting, we perform numerical experiments on two example heating networks. One smaller network, with only two consumers described in detail in \cref{subsec:example}, and one bigger network with 22 consumers for a more realistic and demonstrative result. For the numerical calculations, we used \textit{Julia} with the \textit{NonlinearSolve}\cite{pal2024nonlinearsolve} package to solve \eqref{eq:theorem conditions}.

\subsection{Two-consumer Network} \label{subsec:smallNetwork}
\begin{figure}[h]
    \vspace{0.05cm}
    \centering
    \includegraphics[]{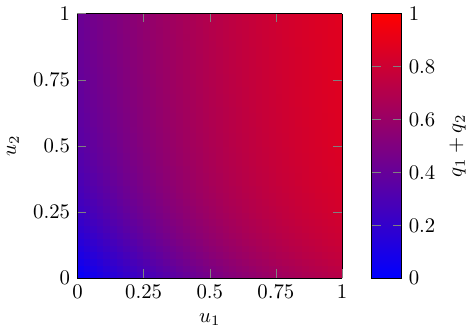}
    \caption{Total flow rates ($q_1 + q_2$) in a 2-consumer network for varying valve positions $u_1$ and $u_2$.}
    \label{fig:2ConsumerTotalFlow}
\end{figure}

\begin{figure}[h]
    \centering
    \includegraphics[]{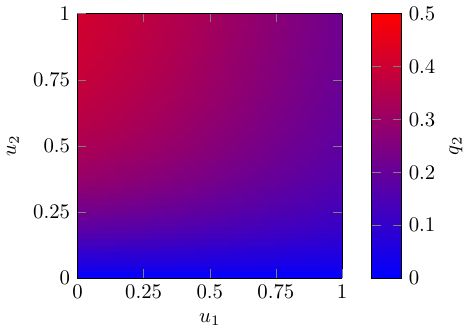}
    \caption{The flow $q_2$, through the second consumer for varying valve positions $u_1$ and $u_2$.}
    \label{fig:2ConsumerQ2Flow}
\end{figure}

For the two consumer network, described in \cref{subsec:example}, we are interested in how the flows change in the system when the valve positions are altered. The theorem states that the total flow through the consumers, in this case given by $q_1+q_2$, should increase whenever either of the valve positions increase. We can see this in \cref{fig:2ConsumerTotalFlow} which shows how the total flow depends on the valve positions $u_1$ and $u_2$. When increasing one valve position without decreasing the other, we see the total flow increase as in \eqref{eq:result 1}. In \cref{fig:2ConsumerQ2Flow} we see a demonstration of the second part of the theorem statement. The figure shows how the flow $q_2$ depends on $u_1$ and $u_2$. For a fixed $u_2$, the flow $q_2$ decreases with increasing $u_1$ as the theorem states. The same is true for the flow $q_1$ when fixing $u_1$ and increasing $u_2$, which can be seen by subtracting the values in \cref{fig:2ConsumerQ2Flow} from \cref{fig:2ConsumerTotalFlow}.
\newpage

\subsection{Large Network} \label{subsec:largeNetwork}

\begin{figure}[t]
    \vspace{.15cm}
    \centering
    \includegraphics[width=0.4\textwidth]{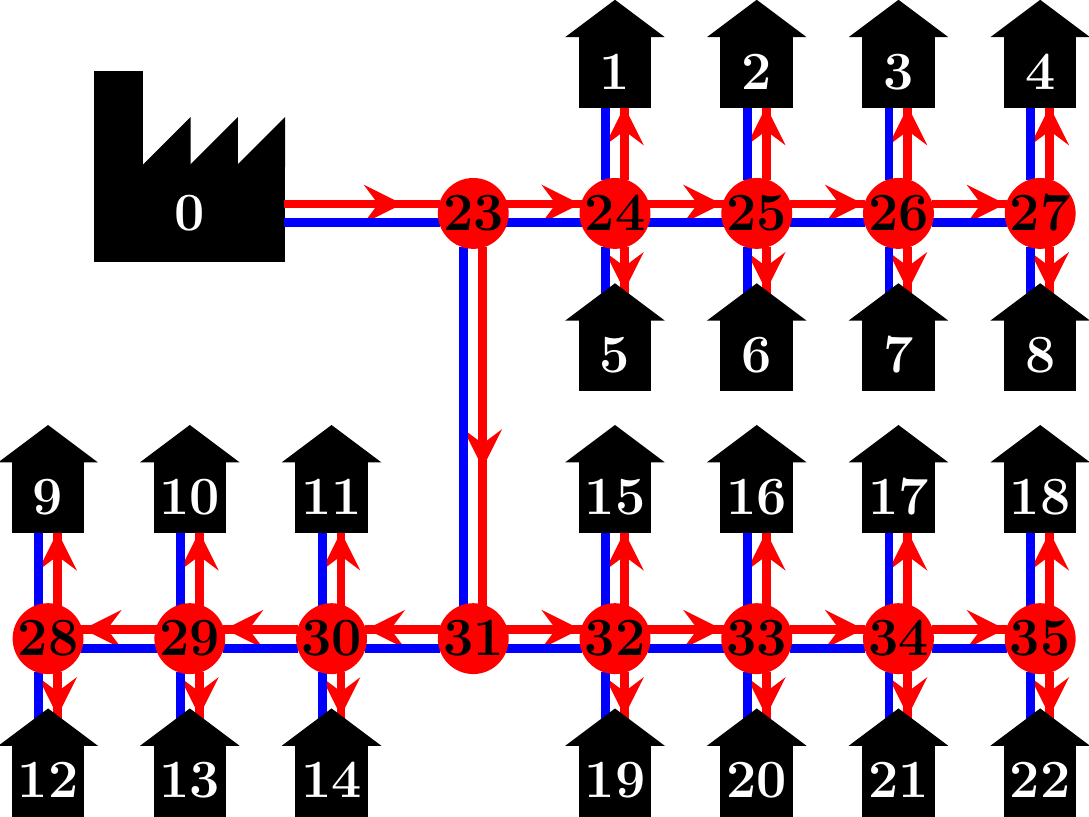}
    \caption{Schematic of a district heating network with 22 consumers.}
    \label{fig:large network}
\end{figure}

\begin{figure}[h]
    \centering
    \includegraphics[]{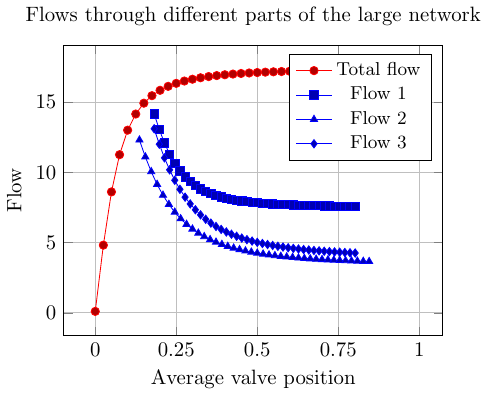}
    \caption{Flow rates through different parts of the system under various valve configurations. The red curve shows the total consumer flow when all valves in the network are opened simultaneously. The blue curves show the flow into each housing group (Flow 1, Flow 2, and Flow 3) when only the valves in the other two groups are opened. For example, Flow 1 represents the flow into Group 1 when only Groups 2 and 3 have open valves.}
    \label{fig:large network flows}
\end{figure}

For the larger, more realistic heating network shown in \cref{fig:large network}, a comprehensive plot showing all dependencies from valve positions to flow rates is infeasible. Instead, we investigate the scenarios described in Theorem 1 by dividing the system into its three main branches: Group 1 (Consumers 1–8), Group 2 (9–14), and Group 3 (15–22).
In \cref{fig:large network flows}, the red curve shows the total flow through the consumers when the valve positions in all houses are opened at the same time. This could realistically occur if a cold front were to hit all locations simultaneously. As predicted by the theorem, the total flow increases. We then examine what happens to each group when only the valves in the other groups are opened. In practice, this phenomenon could result from local temperature differences or varying insulation levels between housing groups. As predicted by the theorem, the group that does not open its valves experiences a drop in flow, as shown by the blue curves in \cref{fig:large network flows}. This means that even if the houses in one of the groups are satisfied with their temperature, they still need to open their valves if the other groups do so in order to maintain their flow.

\section{CONCLUSION} \label{sec:conclusion}
In this paper, we analyzed the hydraulic equilibrium properties of tree-structured district heating networks. We demonstrated that, under mild monotonicity assumptions on the hydraulic components (pumps and valves), key insights can be drawn regarding the mapping from actuator inputs (valve positions) to flow rate distributions in the network. Specifically, we showed that incrementally opening valves guarantees an increase in the total flow to consumers, while keeping a valve closed when other valves are opened results in a decreased flow through that valve. These findings were further supported by numerical simulations conducted on two example networks of different sizes.

\subsection{Future Work}
Future research could explore more general network structures. In particular, extending the analysis to meshed network topologies while maintaining primarily valve-actuated flow rates would provide further insights. Additionally, investigating pump-actuated networks, as considered in \cite{de_persis_output_2014}, could offer valuable perspectives on alternative control mechanisms.

\bibliographystyle{ieeetr}
\bibliography{bibliography}

\end{document}